\documentclass{ws-p8-50x6-00}
\usepackage{amsmath,amssymb}
\newcommand{\ii}{\mathrm{i}}

\newcommand{\dd}{\mathrm{d}}
\newcommand{\pd}{\partial}
\newcommand{\hh}{\mathcal{H}}

\newcommand{\e}{\mathrm{e}}
\newcommand{\ket}[1]{\left|#1\right\rangle}
\newcommand{\bra}[1]{\left\langle #1\right|}

\newcommand{\const}{\mathrm{const}}

\newcommand{\tr}{\mathop{\mathrm{tr}}}

\newcommand{\II}{\mathbb{I}}
\begin{document}

\title{Interacting Noncommutative Lumps}

\author{Corneliu Sochichiu}

\address{Bogoliubov Laboratory of Theoretical
Physics\\
Joint Institute for Nuclear Research\\ 141980 Dubna, Moscow Reg.\\
RUSSIA\\Institutul de Fizic\u a Aplicat\u a A\c S,\\ str. Academiei, nr. 5,\\
MD2028 Chi\c sin\u au,\\ MOLDOVA\\
E-mail: sochichi@thsun1.jinr.ru}

%%%%%%%%%%%%%%%%%%%%%%%%%%%%%%%%%%%%%%%%%%%%%%%%%%%%%%%%%%%%%%
% You may repeat \author \address as often as necessary      %
%%%%%%%%%%%%%%%%%%%%%%%%%%%%%%%%%%%%%%%%%%%%%%%%%%%%%%%%%%%%%%

\maketitle

\abstracts{
We consider interaction of two lumps corresponding to 0-branes in
noncommutative gauge theory}

Recent development of string theory has shown that noncommutative (NC) models
can arise in certain limits of String Theory \cite{Seiberg:1999vs}.

Noncommutative field theory operates with operators defined on a Hilbert
space rather than with ordinary functions. There is, however, a one-to-one
correspondence between such operators and functions which is given by the
Weyl map. Important solutions discovered recently are the so called
noncommutative solitons or lumps \cite{Gopakumar:2000zd}. These
configurations are represented by projector operators. Their Weyl symbols
are localised functions.

In this short note we are going to consider the interaction of two such gauge
field lumps. This is realised by the dynamics of the configuration which
represents a superposition of two projectors to finite-dimensional subspaces
of the Hilbert space which correspond to two copies of the oscillator vacuum
state shifted along noncommutative plane. For more details the reader is
referred to Ref. \cite{Sochichiu:2001am}.
% ----------------------------------------------------------------
\subsection{The Model.} Consider the noncommutative gauge model described by
the following action,
\begin{equation}\label{action}
  S=\int\dd t\tr\left(\frac{1}{2}\dot{X}^i\dot{X}^i+\frac{1}{4g^2}
  [X^i,X^j]^2\right).
\end{equation}
Here fields $X^i$, $i=1,\dots,D$ are time dependent Hermitian operators
defined on some separable infinite-dimensional Hilbert space $\hh$.

In the two-dimensional Moyal--Weyl form some operators $f$ are represented by
functions $f(x)$ on $x^\mu$ $\mu=1,2$, subject to the star-product,
\begin{equation}\label{star}
  f*g(x)=\left.\e^{\ii\theta\epsilon^{\mu\nu}\pd_\mu\pd'_\nu}
  f(x)g(x')\right|_{x'=x},
\end{equation}
where $x^\mu$ satisfy, $[x^1,x^2]=\ii\theta$.

Equations of motion look as follows
\begin{equation}\label{EqM}
  \ddot{X}_i+\frac{1}{g^2}[X_i,[X_i,X_j]]=0.
\end{equation}

In the case of just two lumps one can take them to be:
\begin{subequations}\label{2solit}
\begin{align}\label{2solit1}
  &X_1=cVPV^{-1}\equiv c\ket{-u/2}\bra{-u/2},\\ \label{2solit2}
  &X_2=cV^{-1}PV \equiv c\ket{u/2}\bra{u/2}\\
  &X_i=\const,\qquad i=3,\dots,D,
\end{align}
\end{subequations}
where we introduced the shorthand notations,
\begin{align}\label{V}
  & V=\e^{(\ii/2) p_\mu u^\mu}=
  \e^{\frac{1}{2}(a\bar{u}-\bar{a}u)},\\ \label{P}
  & P=\ket{0}\bra{0}.
\end{align}

To find the time evolution of such a system one has to solve the equations of
motion \eqref{EqM} with initial condition given by \eqref{2solit} and
$\dot{X}_i|_{t=0}=0$.

Symmetries of the model and initial conditions allows one to reduce the
system to a two-dimensional particle described by the equations,
\begin{subequations}\label{2d_part}
\begin{align}\label{X_sc}
  &\ddot{X}=-\frac{2}{g^2}Y^2X, \\ \label{Y_sc}
  &\ddot{Y}=-\frac{2}{g^2}X^2Y,
\end{align}
with the initial conditions,
\begin{equation}\label{ic_XY_sc}
  X|_{t=0}= \e^{-\frac12 |u|^2},\qquad
  Y|_{t=0}=-\sqrt{1-\e^{-|u|^2}}.
\end{equation}
\end{subequations}

This model has a number of interesting features
\cite{Baseian:1979zx,Medvedev:1985ja,Aref'eva:1998mk}. The motion resulting
from above equations appears to be \emph{stochastic\/} for all values of $u$
except $u=\sqrt{\theta}\ln2$. When $u=\sqrt{\theta}\ln2$ the motion is
periodic but unstable, infinitesimal deviation from this position brings it
back to stochasticity.

The solution to equations \eqref{EqM} is expressed in terms linear
combination of some functions $\sigma_a(\bar{z},z)$ with coefficients
depending on $X(t)$ and $Y(t)$,
\begin{align}
  & X_1(t,\bar{z},z)=\frac12
  \sigma_0(\bar{z},z)+ X(t)\sigma_3(\bar{z},z)+
  Y(t)\sigma_1(\bar{z},z),\\
  & X_2(t,\bar{z},z)=\frac12
  \sigma_0(\bar{z},z)+ X(t)\sigma_3(\bar{z},z)-
  Y(t)\sigma_1(\bar{z},z).
\end{align}
Functions $\sigma_a(\bar{z},z)$ are the Weyl symbols of the two-dimensional
Pauli matrices,
\begin{subequations}\label{sigmas}
\begin{align}\label{sigma1}
  \sigma_1(z,\bar{z})&=\frac{2}{\sqrt{1-\e^{-|u|^2}}}
  \left(\e^{-2|z-\frac{u}{2}|^2}-\e^{-2|z+\frac{u}{2}|^2}\right),\\
\label{sigma2}
  \sigma_2(z,\bar{z})&=\frac{2\ii\e^{-2\bar{z}z}}{\sqrt{1-\e^{-|u|^2}}}
  \left(\e^{\bar{z}u-\bar{u}z}-\e^{-\bar{z}u+\bar{u}z}\right),\\
\label{sigma3}
  \sigma_3(z,\bar{z})&=-\frac{2\e^{-\frac12|u|^2}}{1-\e^{-|u|^2}}
  \left(\e^{-2|z-\frac{u}{2}|^2}+\e^{-2|z+\frac{u}{2}|^2}\right)\\
  \nonumber
  & \qquad +\frac{\e^{-2\bar{z}z}}{1-\e^{-|u|^2}}
  \left(\e^{\bar{z}u-\bar{u}z}+\e^{-\bar{z}u+\bar{u}z}\right),\\
\label{sigm0}
  \II(\bar{z},z)&=\sigma_0(z,\bar{z})=
  \frac{2}{1-\e^{-|u|^2}}
  \left(\e^{-2|z-\frac{u}{2}|^2}+\e^{-2|z+\frac{u}{2}|^2}\right)\\
  \nonumber
  & \qquad -\frac{2\e^{-2|z|^2-\frac12 |u|^2}}{1-\e^{-|u|^2}}
  \left(\e^{\bar{z}u-\bar{u}z}+\e^{-\bar{z}u+\bar{u}z}\right).
\end{align}
\end{subequations}

The solution describes lumps bouncing (stochastically) around ``points''
$z=0,\pm u/2$ of the noncommutative plane. In the string theory language the
heights of the lumps can be interpreted as transversal (to the noncommutative
2-brane) coordinates of two 0-branes.

\section*{Acknowledgments}
This work was supported by RFBR grant \#99-01-00190, INTAS grants \#00 262,
Scientific School support grant \# 00-15-96046.

%% ----------------------------------------------------------------
% \bibliographystyle{utphys}
% \bibliography{noncom}
%% ----------------------------------------------------------------

\end{document}